\documentclass[Afour,sageh,times]{sagej}

\usepackage[hidelinks]{hyperref}
\usepackage{graphicx}
\graphicspath{{./images/}}
\DeclareGraphicsExtensions{.pdf,.png,.jpg}

\usepackage{amsmath}
\usepackage{amssymb}
\usepackage{mathtools}
\usepackage{algorithm}
\usepackage{algpseudocode}
\usepackage{optidef}

\usepackage[dvipsnames]{xcolor}
\usepackage{tabularray} 
\usepackage{multirow}
\usepackage{tabularx}
\usepackage{array}
\usepackage{float}
\usepackage{placeins}
\usepackage{upgreek}
\usepackage{moreverb,url}

\newcolumntype{P}[1]{>{\centering\arraybackslash}m{#1}}

\begin{document}

\title{Soft Semi-active Back Support Device with Adaptive Force Profiles using Variable-elastic Actuation and Weight Feedback}

\author{Rohan Khatavkar\affilnum{1}, The Bach Nguyen\affilnum{1}, Inseung Kang\affilnum{2}, Hyunglae Lee\affilnum{1}, and Jiefeng Sun\affilnum{1}}

\affiliation{\affilnum{1}Arizona State University\\
\affilnum{2}Carnegie Mellon University}

\corrauth{Jiefeng Sun}
\email{Jiefeng.Sun@asu.edu}

\begin{abstract}
Portable active back support devices (BSDs) offer tunable assistance but are often bulky and heavy, limiting their usability. In contrast, passive BSDs are lightweight and compact but lack the ability to adapt their assistance to different back movements. We present a soft, lightweight, and compact BSD that combines a variable-stiffness passive element and an active element (an artificial muscle) in parallel. The device provides tunable assistance through discrete changes in stiffness values and active force levels. We validate the device's tuning capabilities through bench testing and on-body characterization. Further, we use the device's tuning capabilities to provide weight-adaptive object lifting and lowering assistance. We detect the weight handled by the user based on forearm force myography and upper-back inertial measurement unit data. Furthermore, electromyography analyses in five participants performing symmetric object lifting and lowering tasks showed reductions in back extensor activity. Preliminary results in one participant also indicated reduced muscle activity during asymmetric lifting.
\end{abstract}

\keywords{Back support device, tunable stiffness, tunable slack, soft wearable robots}

\maketitle


\section{Introduction}

\subsection{Motivation}

Back injuries are among the most common musculoskeletal injuries (\cite{ferguson_prevalence_2019}), with medical expenses in the United States exceeding \$365 billion in 2019 (\cite{lo_systematic_2021}). Back Support devices (BSDs) have the potential to reduce the risk of overexertion, a leading factor contributing to back injuries (\cite{kermavnar_effects_2021}). In addition to alleviating overexertion-related strain, BSDs may also assist individuals with diminished trunk muscle strength, such as older adults, in performing daily activities (\cite{granacher_importance_2013}).

\begin{figure}[h!]
\centerline{\includegraphics[width = 1\linewidth]{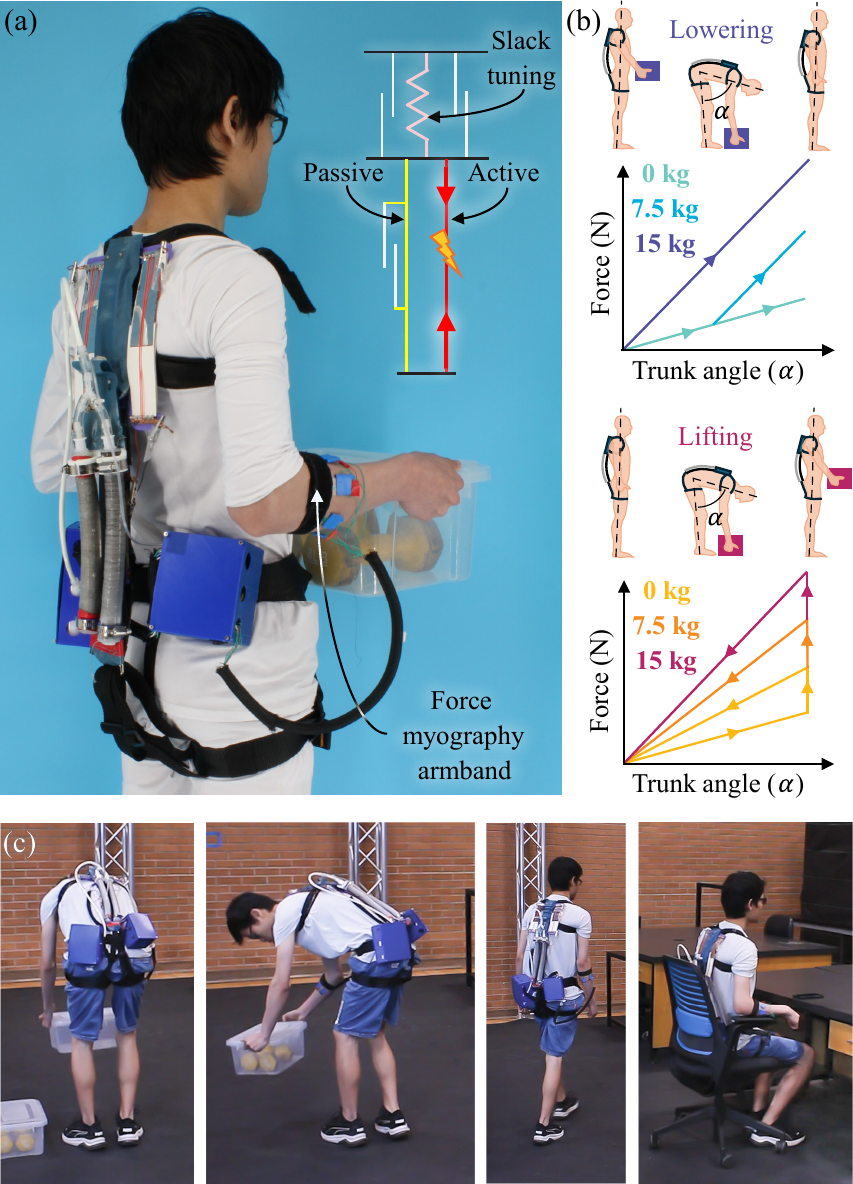}}
\caption{Overview. (a) Our proposed back support device combines a variable-stiffness passive element and an active element (a pneumatic artificial muscle) in parallel. (b) The device implements weight-adaptive assistance using discrete passive and active force profiles. (c) The lightweight and compact soft structure enables assistance during lifting and lowering, while allowing unrestricted mobility during activities such as sitting and walking.}
\label{fig: First Figure}
\end{figure}


BSDs intended for occupational or daily living use must balance wearability with biomechanical efficacy. A lightweight design with a compact form factor and low encumbrance is essential to user acceptance and long-term adoption \cite{roveda_user-centered_2022}. Beyond wearability, operationally useful BSDs must offer task-dependent, tunable assistance \cite{moya-esteban_soft_2025}. Fixed-profile assistance can reduce muscle activity in controlled experiments but fails to accommodate variability in external loads and lifting techniques in real life. To our knowledge, no existing BSD can provide tunable assistance while maintaining a lightweight, compact design. 

\subsection{Related Work}
Based on actuation, state-of-the-art BSDs can be categorized into two main types: passive and active. Passive devices are lightweight and compact, but they do not allow users to adjust the assistive force profile. Although they have been shown to reduce back extensor activity and spinal compression (\cite{lamers_feasibility_2018, koopman_biomechanical_2020, moon_lower-back_2022, refai_benchmarking_2024, schmalz_passive_2022, lamers_design_2021, slaughter_design_2025}), their lack of tunability limits their adaptability. Tuning of the force profile is essential, as optimal force profiles vary substantially across back motions and weights handled by the user (\cite{matijevich_promising_2021, abdoli-e_-body_2006}). Most passive devices use fixed-stiffness elements such as elastic bands, metallic beams, or compliant mechanisms to store and return energy over a trunk flexion and extension cycle (\cite{lamers_low-profile_2020, naf_passive_2018, noauthor_laevo_2024}). Although recent tunable passive devices introduce stiffness scaling mechanisms (\cite{khatavkar_hybrid_2025, ide_evaluation_2021, so_biomechanical_2022, song_development_2024}), tuning must be done before movement begins and remains limited to simple scaling of linear force profiles as a function of the trunk flexion angle. Furthermore, passive devices rely on the user to preload the device during trunk flexion. This implies that higher assistive forces require higher stiffness values. This can restrict the user's natural motion and comfort (\cite{baltrusch_testing_2020, poliero_versatile_2021, baldassarre_industrial_2022}). Additionally, passive devices often exhibit hysteresis, which potentially implies a higher energetic burden on the user when using the device compared to not using it (\cite{van2022measuring}). In summary, passive devices are lightweight and compact, but their assistance cannot be tuned, and they cannot perform net positive work on the user's body.

\begin{table*}[t]
\small\sf\centering
\caption{Comparison of existing untethered BSDs.}
\begin{tabular}{cP{1.25cm}P{1.25cm}P{2.75
cm}P{0.8cm}P{0.8cm}P{1
cm}P{1.4cm}P{1cm}}
\toprule
\textbf{Device} & \textbf{Active or Passive} & \textbf{Rigid or Soft} & \textbf{Force Generating Element} & \textbf{Weight (kg)} & \textbf{Peak Torque (Nm)} & \textbf{Torque Density (Nm/kg) } & \textbf{Weight Adaptive Profile} & \textbf{EMG Reduction (\%)}\\
\hline
\cite{slaughter_design_2025} & Passive & Soft & Elastic band & 1.4 & 35 & 25 & No & NA\\
\hline
\cite{koopman_effects_2020} & Passive & Rigid & Compliant mechanism & 4 & 30 & 7.5 & No & 8\\
\hline
\cite{song_development_2024} & Passive & Rigid & Compliant mechanism & 5 & 7.04 & 1.41 & No & NA\\
\hline
\cite{khatavkar_hybrid_2025} & Passive & Soft & Variable stiffness elastic band & 1.2 & 20.4 & 17 & No & NA\\
\hline
\cite{lanotte_adaptive_2021} & Active & Rigid & Direct drive motor & 8 & 35 & 4.375 & No & 36.3\\
\hline
\cite{li_development_2023} & Active & Rigid & Direct drive motor & 6 & N/A & N/A & No & 11.63\\
\hline
\cite{poliero_active_2022} & Active & Rigid & Direct drive motor & 8 & 40 & 5 & No & 41\\
\hline
\cite{liao_design_2024} & Active & Rigid & SEA & 6.5 & 75 & 11.54 & No & 22.7\\
\hline
\cite{ding_novel_2024} & Active & Rigid & SEA & 5 & 70 & 14 & No & 12.51\\
\hline
\cite{hyun_singular_2020} & Active & Rigid & SEA & 5.5 & 94.5 & 17.18 & No & 33\\
\hline
\cite{heo_semg-triggered_2022} & Active & Rigid & Pneumatic cylinder & 9.2 & 80 & 8.7 & No & 25.1\\
\hline
\cite{song_cable-driven_2023} & Active & Rigid & Motorized spine & 6.3 & 88.2 & 14 & No & 43.75\\
\hline
\cite{in_kim_bilateral_2024} & Active & Rigid & Motorized spine & 5.75 & N/A & N/A & No & NA\\
\hline
\cite{chen_development_2025} & Active & Rigid & Motorized spine & 4.83 & 130 & 26.92 & No & 41.28\\
\hline
\cite{cullen_reducing_2026} & Active & Soft & Cable-drive & 4.35 & 76 & 17.47 & No & 6.7\\
\hline
\cite{chung_lightweight_2024} & Active & Soft & Cable-drive & 2.7 & 30 & 11.1 & No & 17.6\\
\hline
\textbf{This Device} & \textbf{Semi-active} & \textbf{Soft} & \textbf{Variable-elastic actuator} & \textbf{1.97} & \textbf{29} & \textbf{14.76} & \textbf{Yes} & \textbf{15}\\
\hline
\bottomrule
\end{tabular}\\[10pt]
\label{table:g}
\end{table*}

Active BSDs can tune assistive force profiles and perform significant net positive work on the user’s body, but they are heavy and bulky, which limits their wearability. They use active elements like electric motors to generate the assistive trunk torque (\cite{toxiri_parallel-elastic_2018, chung_lightweight_2024, zhu_smart_2023}). Many active BSDs actuate linkages with a revolute joint that is aligned with the biological hip joint using direct-drive motors (\cite{lanotte_adaptive_2021, li_development_2023, poliero_active_2022}), series elastic actuators (SEAs) (\cite{liao_design_2024, ding_novel_2024, hyun_singular_2020}), or pneumatic cylinders (\cite{heo_backdrivable_2020, heo_semg-triggered_2022}). Although these devices offer assistive force-tuning capabilities and demonstrate a significant reduction in back extensor effort, they are invariably heavy and bulky, and unsuitable for daily use. Active BSDs, which incorporate an underactuated high-degree-of-freedom spine in parallel with the biological spine, have demonstrated reduced back extensor effort and reduced spinal compression. However, they are either tethered or become bulky when untethered (\cite{yang_spine-inspired_2019, song_cable-driven_2023, in_kim_bilateral_2024, chen_development_2025}). Cable-driven devices that generate tension in a cable connecting the user's upper back to the thigh interfaces (\cite{li_design_2022, chung_lightweight_2024}) have also shown a significant reduction in back extensor effort. However, cable-driven devices also use heavy and bulky actuation units. Ultimately, while active devices offer desirable force profile tuning, their form factor and weight continue to limit real-world use.

In addition, force profile tuning alone cannot be efficacious in active devices without reliable user state estimation, 
like detection of trunk motion and the handled weight (\cite{babic_challenges_2021}). However, existing automated weight detection can only be implemented in industrial settings with color-coded weights or with off-board computing. Recent studies have explored estimation of the weight handled by the user using computer vision and electromyography (EMG) (\cite{chung_perceptual_2025, moya-esteban_soft_2025, jiang_loading_2024}). However, camera-based approaches are suitable only for handling color-coded objects, such as in industrial settings. EMG-based methods are sensitive to electrode placement and signal quality. They also require high sampling rates and complex electronics, which limit their portability. Thus, the challenging state estimation further limits real-world use of state-of-the-art active devices in daily life.

Further, state-of-the-art BSDs can also be classified as soft or rigid devices. Rigid devices that use rigid linkages with a revolute joint at the user's hip often restrict the user's RoM or use bulky and complex mechanisms to provide transparency (\cite{naf_passive_2018, ding_novel_2024}). Additionally, misalignment between the device's joints and the biological joints can result in unwanted forces, leading to discomfort. BSDs with a high degree of freedom underactuated spine in parallel with the biological spine mitigate the problem of RoM restriction and joint misalignment to only a limited extent (\cite{yang_spine-inspired_2019, song_cable-driven_2023, in_kim_bilateral_2024, chen_development_2025}). Soft BSDs have the potential to resolve the RoM restriction and joint misalignment problems. However, most state-of-the-art soft BSDs are passive and do not provide a tunable force, as soft active elements often cannot provide adequate force when untethered~(\cite{di_lallo_untethered_2024}).

\subsection{Contributions}
In this paper, we present an untethered, autonomous soft BSD that provides adaptive assistive force in a lightweight and compact design. The device achieved torque density and back extensor EMG reduction comparable to significantly heavier active devices, as shown in Table \ref{table:g}. 
The major contribution of the work is striking a balance between passive devices and active devices, via a variable-elastic actuator, as shown in Fig. \ref{fig: First Figure}(a). Our variable-elastic actuator combines a variable-stiffness passive element in parallel with an active element (a pneumatic artificial muscle). Although a large active element alone can generate the entire required force, such a design would inevitably become bulky and heavy. 
The automatic selection of the assistance force profile (see Figure \ref{fig: First Figure}(b) and (c)) is facilitated by on-board state and weight detection. This is achieved by combining forearm force myography (FMG) and inertial measurement unit (IMU) data collected from the upper back.


The remainder of the paper is organized as follows: Sec. \ref{sec:II} describes the device design; in Sec. \ref{Sec:Model} we describe analytical modeling of our device; in Sec. \ref{sec:Dev Char}, we present the bench tests to validate the device design and the analytical model; Sec. \ref{Sec:Moco} proposes a coupled device-musculoskeletal model based force profile optimization for lifting and lowering tasks; Sec. \ref{Sec:SM} presents the automated user state and weight classifiers used to implement weight adaptive assistance; in Sec. \ref{sec:human exp}, we validate the tuning capabilities of our device with on-body characterization and show the biomechanical benefit of the device through EMG experiments; Sec. \ref{sec:conclusion} is the conclusion.

\section{Device Design} \label{sec:II}

\begin{figure}[b!]
\centerline{\includegraphics[width = 1\linewidth]{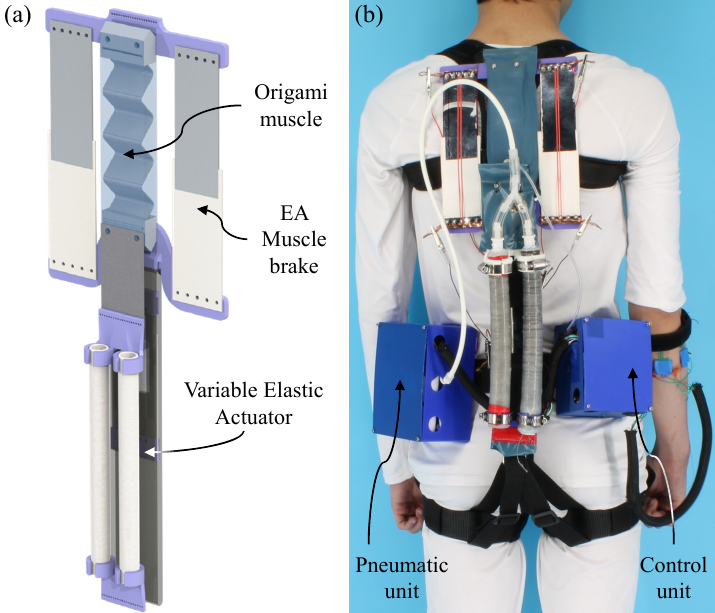}}
\caption{(a) Device assembly with the VS resistance band (passive), IPAMs (active), and origami muscle (slack tuning). (b) Device and portable pneumatic/electronic units worn by a user.}
\label{fig:overall_design}
\end{figure}

The proposed BSD consists of two key components: a variable-elastic actuator and a slack-tuning assembly arranged in series, as shown in Fig.~\ref{fig:overall_design}.
The variable-elastic actuator provides adaptive assistance force, and the slack-tuning assembly can change the device's slack.  
 

\subsection{Variable-Elastic Actuator}\label{sec:variable-stiffness}
VEA combines a variable-stiffness (VS) resistance band and an inverse pneumatic artificial muscle (IPAM) in parallel.

\subsubsection{Design}
\begin{figure}[b!]
\centerline{\includegraphics[width = 1\linewidth]{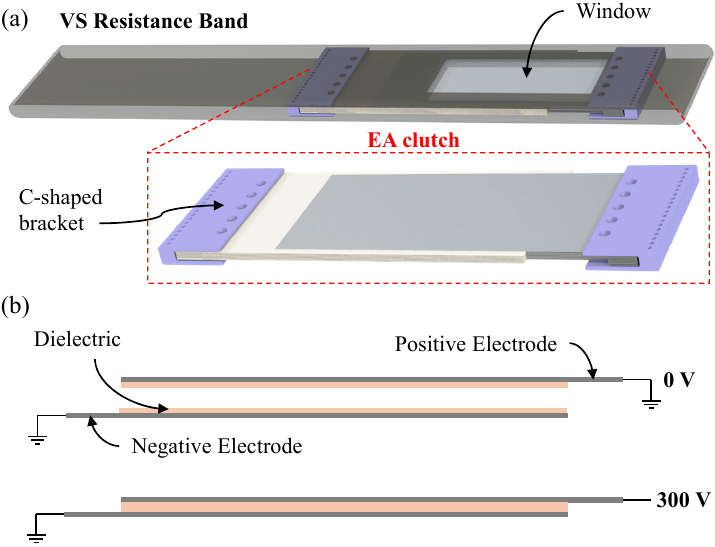}}
\caption{(a) Design of the VS resistance band with an integrated electroadhesive (EA) clutch. (b) Principle of operation of a single interface of the EA clutch: When 300 V is applied between the positive and negative electrodes, the electro-adhesion at the interface prevents sliding. This reduces the active length of the resistance band and increases stiffness. The EA clutch includes eight such interfaces.}
\label{fig:res_band}
\end{figure}

The VS resistance band dynamically adjusts its stiffness to supplement the force of the IPAM. The stiffness change is realized by altering the band’s effective length using an electroadhesive (EA) clutch integrated in parallel with an elastic band, as illustrated in Fig.~\ref{fig:res_band}(a). The stiffness tuning enables weight‑adaptive assistance during trunk flexion. When heavier loads are detected, the EA clutch is engaged earlier in the RoM to increase passive force, as shown in Fig.~\ref{fig: First Figure}(b). Adjusting stiffness rather than increasing IPAM force is energy efficient: the EA clutch draws only 1\,mA during its 300\,ms engagement period and negligible current thereafter. The EA clutch can switch states within 300\,ms, enabling stiffness tuning during a back motion. To ensure rapid release, a 300\,V AC signal at 10\,Hz is applied during disengagement to prevent charge accumulation at the electrode interfaces.

The EA clutch consists of eight positive and eight negative interleaved electrodes, forming eight electroadhesive interfaces. Each electrode pair is mounted to the resistance band via C‑shaped brackets. In the \emph{disengaged} state (0\,V), the electrodes slide freely over each other. When a 300\,V DC potential is applied, the electrodes adhere at their interfaces (\emph{engaged} state), effectively preventing sliding (Fig.~\ref{fig:res_band}(b)). When engaged, the segment of the resistance band parallel to the EA clutch becomes effectively inextensible, thereby increasing the overall stiffness. A window cut into the band between the C‑shaped brackets further amplifies the stiffness change between the engaged and disengaged states.

Two IPAMs are integrated in parallel with the VS resistance band, as shown in Fig.~\ref{fig:overall_design}(a). The IPAM is a soft actuator that elongates upon pressurization and contracts upon deflation~(\cite{hawkes_design_2016}). 
In our device, the IPAM is inflated during trunk flexion, ensuring that only the VS resistance band applies a force and stores energy. At the onset of trunk extension, the IPAM is rapidly deflated to generate additional assistive force that augments the energy returned by the VS resistance band. By adjusting the deflation pressure, the active force can be tuned as desired. Heavier weights result in greater IPAM deflation, based on weight detection from forearm FMG signal (Fig.~\ref{fig: First Figure}(b)).


\subsubsection{Fabrication}

The VS resistance band was fabricated from a 300\,mm $\times$ 50\,mm elastic band (HCT Style). A 60\,mm $\times$ 30\,mm through‑hole was cut to create the window. The top end of the elastic band was sewn to the lower muscle mount, and the bottom end was attached to the thigh interface of a climbing harness (Fusion, TCH‑107‑2139‑ES‑S‑XL‑BLK‑ORG) using inextensible fabric straps, as shown in Fig.~\ref{fig:overall_design}(b).

Fabrication of the EA clutch followed the method of Diller \textit{et al.}~(\cite{diller_effects_2018}). Each electrode was made from aluminized PET film (McMaster‑Carr \#7538T12). A high‑permittivity dielectric ink (114A, Creative Materials Inc.) was applied to the aluminized surface using a thin‑film applicator to form a 30\,$\mu$m dielectric layer. The coated PET film was cured at 130$^\circ$C under vacuum for two hours, left at room temperature overnight, and vacuumed again at 130$^\circ$C for two hours. The coated PET film was then cut to size to form individual electrodes. 
Electrical connections were established by sewing a 30\,AWG copper wire onto an exposed aluminum region of each electrode and reinforcing it with copper tape to ensure low‑resistance contact. The electrodes were punched, stacked, and bolted to the C‑shaped brackets. Finally, the C‑shaped brackets with the electrodes were sewn into the elastic band to form the complete VS resistance band.


\begin{figure}[b!]
\centerline{\includegraphics[width = 1\linewidth]{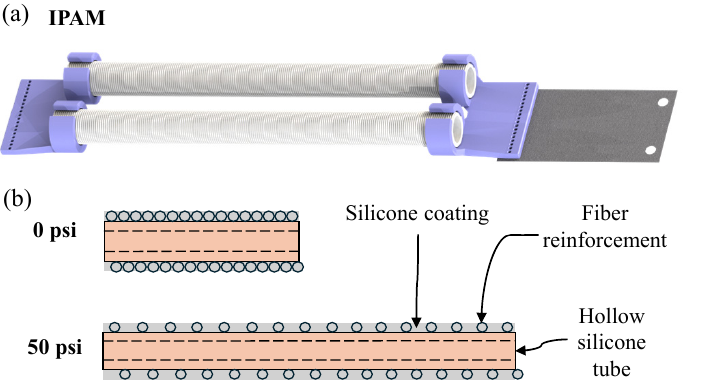}}
\caption{(a) IPAM design: It consists of a silicone tube helically wound with fiber reinforcement and sealed at both ends. Our device uses two such IPAMs in parallel. (b) Working principle of the IPAM: When pressurized, it extends axially; when deflated, it contracts, thereby generating an active force.}
\label{fig:IPAM}
\end{figure}

Each IPAM was fabricated by molding a custom silicone tube (200\,mm length, 12.5\,mm inner diameter, 25\,mm outer diameter) using Dragon Skin 30 (SmoothOn). One end of the tube was sealed with a barbed plug, while the other was fitted with a straight barbed connector to form the air inlet. The tube was then helically wound with nylon thread, ensuring zero spacing between adjacent turns. Finally, the entire assembly was coated with a thin layer of silicone rubber to encapsulate the fibers and prevent slippage during repeated pressurization and deflation.

\subsection{Slack Tuning Assembly}

\begin{figure}[b!]
\centerline{\includegraphics[width = 1\linewidth]{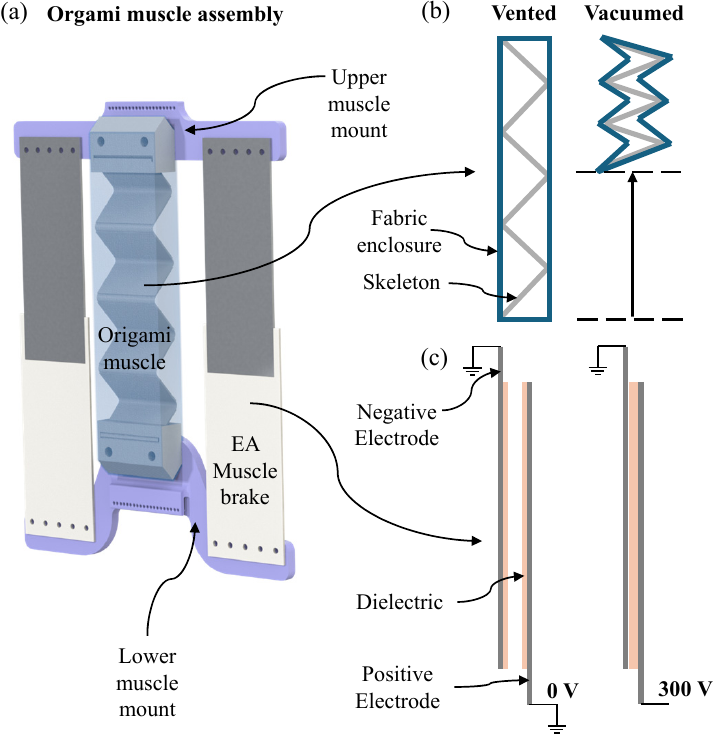}}
\caption{(a) Origami muscle assembly for slack tuning. (b) Working principle of the origami muscle: On vacuuming, a zigzag skeleton housed in a sealed fabric enclosure contracts, shortening the device. (c) Working principle of the EA muscle brakes: When 300 V is applied between the positive and negative electrodes, the electro-adhesion at the interface prevents sliding, locking the origami muscle length after slack tuning. This enables compact actuation without constant power draw. The EA muscle brakes include five EA interfaces on either side of the origami muscle.}
\label{fig:Origami_muscle_brake}
\end{figure}

A vacuum-driven origami muscle is used to tune slack in the device, as shown in Fig.~\ref{fig:Origami_muscle_brake}(a). This actuator offers a high contraction ratio, which enables a wide slack adjustment range~(\cite{li_fluid-driven_2017}). It consists of a zig-zag folded skeleton sealed in a fabric enclosure. When vacuum is applied, the fabric enclosure collapses inward, causing the skeleton to fold and contract. Releasing the vacuum allows the skeleton to passively return to its rest length (Fig.~\ref{fig:Origami_muscle_brake}(b)).

To fabricate the origami muscle, first we 3D-printed the zig-zag skeleton using polylactic acid (PLA) on a Prusa MK4 printer. The skeleton is 40\,mm wide. Then, the skeleton was enclosed in heat-sealable, polyurethane-coated nylon fabric (Seattle Fabrics), which was then heat-sealed to form an airtight enclosure. The completed muscle assembly was bolted to the upper and lower muscle mounts.

To hold the origami muscle at a desired slack length, we incorporated electroadhesive brakes on both sides of the muscle, as shown in Fig.~\ref{fig:Origami_muscle_brake}(a). These brakes (denoted by EA muscle brakes) operate on the same principle as the EA clutch as shown in Fig. \ref{fig:Origami_muscle_brake}(c) and described in Sec.~\ref{sec:variable-stiffness}. Each EA muscle brake consists of five interleaved pairs of positive and negative electrodes mounted between the upper and lower muscle mounts.
The EA muscle brakes remain disengaged during origami muscle actuation to allow slack adjustment and are engaged once the desired slack length is reached. This braking mechanism serves two key purposes. First, it allows the use of a compact origami muscle, which can be feasibly actuated using a portable pneumatic system. Without the brake, a much larger muscle would be required to generate the force needed to hold the slack length under load. Second, the EA muscle brakes are energy efficient as they require power only during engagement and consume negligible power once engaged, which contributes to improved battery life. Electrode fabrication for the EA muscle brakes followed the same procedure used for the VS resistance band, as detailed in Sec.~\ref{sec:variable-stiffness}.

\subsection{FMG Armband}

\subsubsection{Design}

The FMG armband is designed to measure the radial force exerted by forearm muscles as they expand during object lifting. Since this radial expansion scales with the weight being handled, the FMG signal can be used to infer lifting effort and enable weight-adaptive assistance~(\cite{sakr2019estimation,wininger2008pressure}).

\begin{figure}[b!]
\centerline{\includegraphics[width = 1\linewidth]{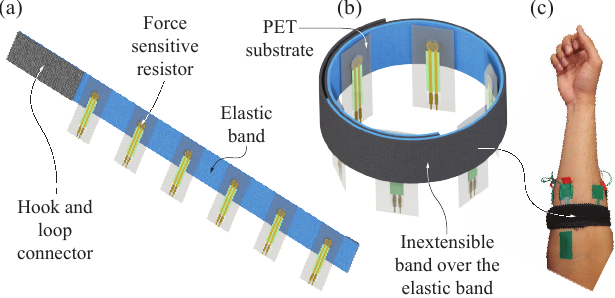}}
\caption{(a) FMG armband design: Six FSRs mounted on PET substrates measure radial forearm pressure during weight handling. (b) An inextensible band wrapped over the elastic band limits excessive radial expansion of the elastic band when the user handles a weight. (c) FMG armband worn by a human subject.}
\label{fig:armband}
\end{figure}

The armband incorporates six force-sensitive resistors (FSRs) (FSR400, Interlink Electronics) to capture the radial pressure on the forearm. Each FSR is mounted on a polyethylene terephthalate (PET) substrate that provides a rigid backing for consistent force measurement. These PET-backed sensors are sewn onto an elastic band as shown in Fig.~\ref{fig:armband}(a), allowing the armband to conform to a wide range of forearm sizes. The elastic band also ensures that the sensors remain equi-spaced regardless of the user's forearm size. To further enhance measurement reliability, an inextensible band is wrapped over the elastic band and secured using a hook-and-loop connector, as illustrated in Fig.~\ref{fig:armband}(b). This inextensible band limits excessive radial expansion of the elastic band when the user handles a weight, enabling reliable detection of subtle changes in radial pressure. Fig.~\ref{fig:armband}(c) shows the fully assembled armband worn by a user.

\subsubsection{Fabrication}

Six PET substrates (12\,mm $\times$ 40\,mm $\times$ 0.254\,mm) were first sewn at equal spacing along a 300\,mm long elastic band. The band length was selected based on the average forearm circumference at 15 \% of forearm length from the elbow of the participants in our human experiments. FSRs were adhered to the PET substrates using double-sided adhesive. Hook-and-loop connectors were sewn to opposite ends of the elastic band to form an adjustable loop that can be slid onto the user's forearm. An inextensible band was then wrapped around the elastic band and affixed using a hook-and-loop connector. Finally, the FSRs were connected to the electronic control system.

\begin{figure}[b]
\centerline{\includegraphics[width = 1\linewidth]{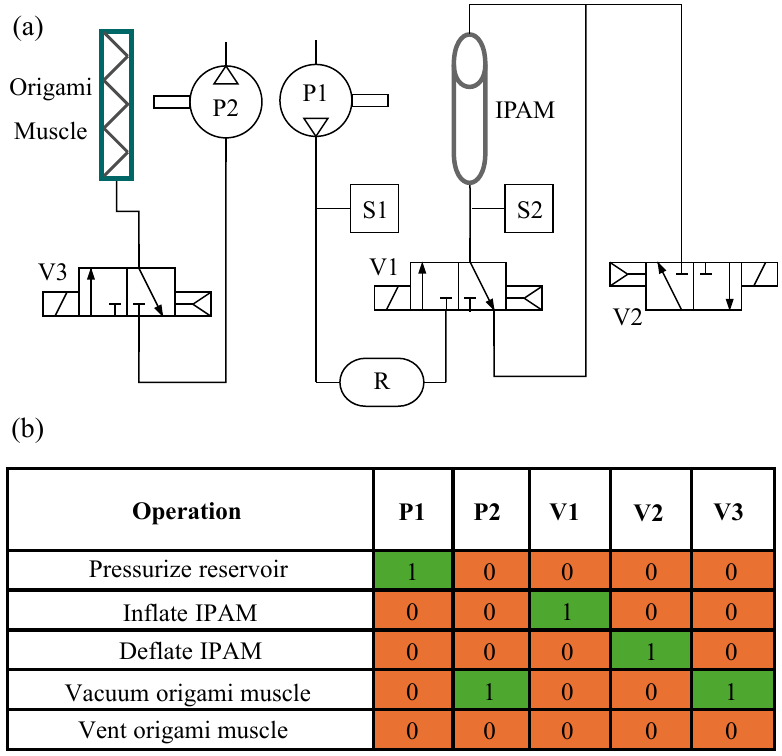}}
\caption{(a) Pneumatic system schematic. (b) Truth table depicting the pneumatic control logic.}
\label{fig:pneumatic_circuit}
\end{figure}

\begin{figure}[b]
\centerline{\includegraphics[width = 1\linewidth]{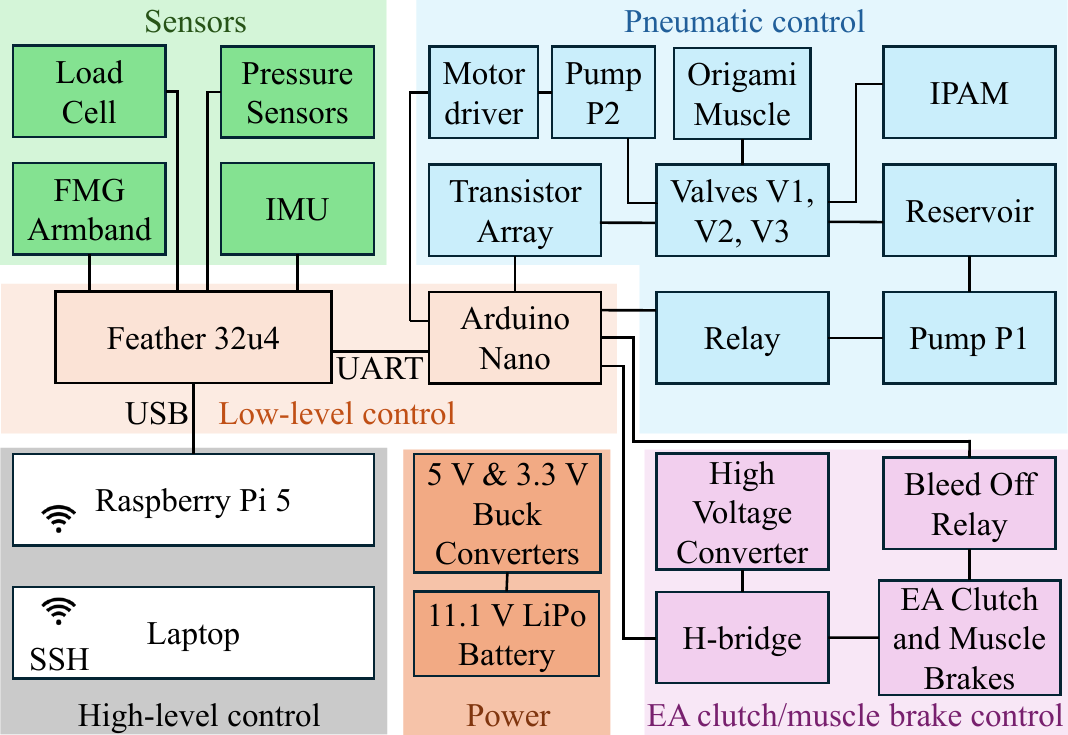}}
\caption{Schematic of the electronic system used to drive the EA clutch, the IPAMs, and the origami muscle assembly.}
\label{fig:elec_circuit}
\end{figure}

\subsection{Pneumatic and Electronic Systems}\label{sec:pneumatics and electronics}

\subsubsection{Pneumatic System}

The device is actuated by a compact, lightweight pneumatic system designed for portable operation. As shown in Fig.~\ref{fig:pneumatic_circuit}(a), a piston pump (P1, BD-07A38L-12V, Bodenflo; 455\,g, 75\,dB) pressurizes three air reservoirs (R, 115.53\,cc each; x-small, Robart), which serve as the compressed air source for the IPAMs. Air from the reservoirs is routed through a 3-port, 2-way solenoid valve (V1, VQ110U-6M, SMC) to the IPAM. A second solenoid valve (V2, VQ110U-6M, SMC) is connected to both the IPAM and the exhaust port of V1 to either hold or release the IPAM pressure. The direct connection between V2 and the IPAM enables rapid deflation. Two pressure sensors (S1 and S2, 4525-SS5A100GP, TE Connectivity) are also included: S1 measures reservoir pressure to maintain it at 90\,psi, while S2 measures IPAM pressure for bang-bang pressure control. To actuate the origami muscle, a micro-compressor (P2, 2000 series, Dynaflo; 20\,g, 65\,dB) draws vacuum through a miniature 3-port, 2-way piezoelectric valve (V3, x-valve, Parker). A truth table summarizing the pneumatic control logic is shown in Fig.~\ref{fig:pneumatic_circuit}(b).

\subsubsection{Electronic Control System}

A schematic of the electronic control system is shown in Fig.~\ref{fig:elec_circuit}. The system uses a primary microcontroller (Feather 32u4, Adafruit) to acquire sensor data and transmit it to a single-board computer (Raspberry Pi 5) via Universal Serial Bus (USB) communication. The Raspberry Pi runs a machine learning algorithm described in Sec.~\ref{Sec:SM} to classify the user's state and the weight being handled. These classification outputs are sent back to the Feather board, which determines the control strategy and generates corresponding actuation commands. A secondary microcontroller (Arduino Nano) receives these commands via Universal Asynchronous Receiver-Transmitter (UART) communication and actuates the pneumatic components accordingly. The piston pump (P1) is controlled via an electromechanical relay, while the micro compressor (P2) is driven by a motor driver (TB6612FNG, Toshiba). Solenoid valves are switched using a transistor array (TBD62083APG-ND, Toshiba).

To actuate the EA clutches and EA muscle brakes, the system includes a high-voltage converter (UMV0505, HMV Tech) capable of generating up to 500\,V. The high-voltage output is fed to two solid-state H-bridge circuits, one each for the EA clutch and the EA muscle brakes. Each H-bridge consists of p-type MOSFETs (DMP45H150DHE-13, Diodes Inc.) on the high voltage side and n-type MOSFETs (IPN70R1K2P7SATMA1, Infineon Tech) on the low voltage side. Optical isolators (TLP293, Toshiba) ensure electrical isolation between high and low-voltage sides. A bleed-off relay (LCC110, Littlefuse) is used to safely discharge the EA clutch and brake by shorting their terminals after disengagement.

Power for the pneumatic system is provided by an 11.1\,V 5200\,mAh battery, which supplies the piston pump and valves. A 5\,V buck converter powers the micro compressor, the Nano board, the high-voltage converter, and pressure sensors. A 3.3\,V buck converter supplies power to the Feather 32u4, IMU, load cell, and the FMG armband’s FSRs.

Sensor data acquisition is handled as follows: the IMU (BNO055, Adafruit), located on the back attachment, communicates with the Feather board via I\textsuperscript{2}C. Load cell data is captured using a differential analog-to-digital converter (NAU7802, Adafruit), also interfaced over I\textsuperscript{2}C. The load cell was used only during the on-body characterization experiments (Sec.~\ref{sec:human exp}) and is not part of the device. For the FMG armband, the change in the FSR resistance is measured using a voltage divider with a 10\,k$\Omega$ fixed resistor. The six FSR signals are transmitted via a 16:1 analog multiplexer (CD4067BE, Texas Instruments) to the Feather board. All sensor data were sampled at 100\,Hz.

\section{Device Model}\label{Sec:Model}
This section describes an analytical model that facilitates device design for achieving weight adaptive force profiles.

\begin{figure}[t]
\centerline{\includegraphics[width = 1\linewidth]{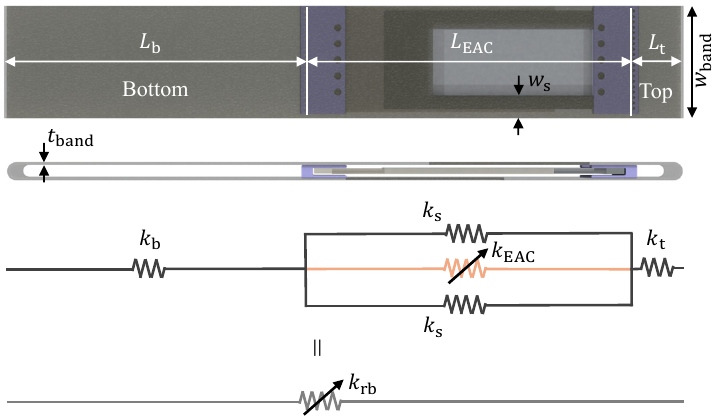}}
\caption{Analytical model of the VS resistance band: Spring-network model used to derive stiffness of the resistance band as a function of geometric and material parameters.}
\label{fig: model}
\end{figure}

\subsection{Variable-stiffness Band Model}
The VS resistance band model was used to select appropriate design parameters to achieve the force profiles for lowering tasks shown in Fig. \ref{fig: First Figure}(b). The design parameters include the dimensions of the resistance band and the EA clutch. The VS resistance band is modeled using an equivalent spring network shown in Fig. \ref{fig: model}.

The VS resistance band's stiffness $k_{\text{rb}}$ is given by
\begin{equation}
k_{\text{rb}}(k_{\text{EAC}}) = \frac{1}{\left( \frac{1}{k_{\text{b}}} + \frac{1}{2k_{\text{s}}+k_{\text{EAC}}} + \frac{1}{k_{\text{t}}} \right)},
\label{eqn:k-rb}
\end{equation}
where $k_{\text{b}}$ and $k_{\text{t}}$ are the stiffness of the bottom and top regions of the VS resistance band, $k_{\text{s}}$ is the stiffness of the VS resistance band regions beside the window, and $k_{\text{EAC}}$ is the stiffness of the EA clutch that changes based on whether it is engaged or disengaged. $k_{\text{b}}$, $k_{\text{t}}$, and $k_{\text{s}}$ are estimated using the formula $
{w_{\text{band}} t_{\text{band}}  E_{\text{band}}}/{L_{\text{band}}},$
where $L_{\text{band}}$ $w_{\text{band}}$, $t_{\text{band}}$ are the length, width, and thickness of the corresponding regions, and $E_{\text{band}}$ is the Young's modulus of the elastic band (determined via tensile testing). $k_{\text{EAC}}$ is the stiffness of the EA clutch given by
\begin{equation}
k_{\text{EAC}} =
\begin{cases}
\infty & \text{if EA clutch is engaged,} \\
0 & \text{if EA clutch is disengaged.}
\end{cases}
\label{eqn:kpi}
\end{equation}
We assumed that $k_{\text{EAC}}$ was sufficiently large compared to $k_{\text{b}}$  and can be treated as $\infty$ when it is engaged.

\subsection{IPAM Dynamics Model}

The IPAM dynamics model guided the design of the IPAM and the pneumatic system. Though static IPAM force models exist in the literature, no dynamic model has been reported. In the subsection, we developed a dynamic model to predict the IPAM force. We assume that the IPAM force dynamics are largely governed by the pressure dynamics, and the IPAM viscoelasticity can be neglected. We validated our assumption by comparing experimentally measured device force with model predictions (\ref{sec:Dev Char}).

Each IPAM is modeled as a cylindrical elastic tube with fixed internal radius $r$ and variable length $L(t)$. The instantaneous internal volume of two IPAMs is given by
\begin{equation}
    V(t) = 2\pi r^2 L(t).
\label{eq:V_dyn}
\end{equation}

The deflation of the IPAM is modeled as compressible, isothermal airflow through a restrictive exhaust line. The internal pressure $P(t)$ decreases over time due to mass loss, causing IPAM contraction.
The initial flow is choked , and transitions to subsonic according to Eqn. \ref{eq:pressure_ratio}:
\begin{equation}
\frac{P_{\text{atm}}}{P(t)} \leq \left( \frac{2}{\gamma + 1} \right)^{\frac{\gamma}{\gamma - 1}},
\label{eq:pressure_ratio}
\end{equation}
where $P(t)$ is the internal pressure of IPAM,  $P_{\text{atm}}$ is the atmospheric pressure and $\gamma$ is the adiabatic index of air.

In our device, the IPAM is initially inflated to $50$ psi gauge, and deflates to atmospheric pressure. Once internal pressure drops below approximately $28$ psi, it transients to a subsonic flow.


When the flow is choked, the mass flow rate is given by:
\begin{equation}
\dot{m}(t) = C_d A_{\text{eff}} P(t) \times\sqrt{ \frac{\gamma}{RT} }
\left( \frac{2}{\gamma+1} \right)^{ \frac{\gamma+1}{2(\gamma - 1)} },
\end{equation}
where $C_d$ is the effective discharge coefficient, $A_{\text{eff}}$ is the effective cross-sectional area of the exhaust line, $R$ is the specific gas constant, and $T$ is the absolute temperature (assumed constant).

When the flow is subsonic:
\begin{multline}
    \dot{m}(t) = C_d A_{\text{eff}} P(t) \\ \times \sqrt{ \frac{2\gamma}{RT(\gamma - 1)} \cdot \bigg[
    \left( \frac{P_{\text{atm}}}{P(t)} \right)^{\frac{2}{\gamma}}
    - \left( \frac{P_{\text{atm}}}{P(t)} \right)^{\frac{\gamma + 1}{\gamma}} \bigg]}.
\label{eq:p_dyn_1}
\end{multline}

$C_d$ and $A_{\text{eff}}$ account for the total resistance to flow in the IPAM deflation path. The IPAM deflation path comprises barbed connectors, elbow and tee joints, a solenoid valve with an internal restriction and a 90° bend, and tubing. We use an equivalent orifice approximation in which all distributed and minor losses are lumped into a single flow restriction. Frictional losses in the tubing are estimated using the Darcy–Weisbach equation, while localized losses due to fittings and bends are accounted for using standard minor loss coefficients. The cumulative pressure loss of these components is incorporated by adjusting the nominal discharge coefficient and the cross-sectional area of the equivalent orifice to reflect the additional flow resistance. These adjusted $C_d$ and $A_{\text{eff}}$ are then used  to compute the mass flow rate $\dot{m}(t)$.

The mass of air in the IPAM is updated at each time step, $\Delta t$, using
\begin{equation}
    m(t+\Delta t) = m(t) - \dot{m}(t)\Delta t,
\end{equation}
and the internal pressure is updated via the ideal gas law:
\begin{equation}
    P(t+\Delta t) = \frac{m(t+\Delta t)RT}{V(t+\Delta t)}.
\label{eq:p_dyn}
\end{equation}

\begin{figure}[t!]
\centerline{\includegraphics[width = 1\linewidth]{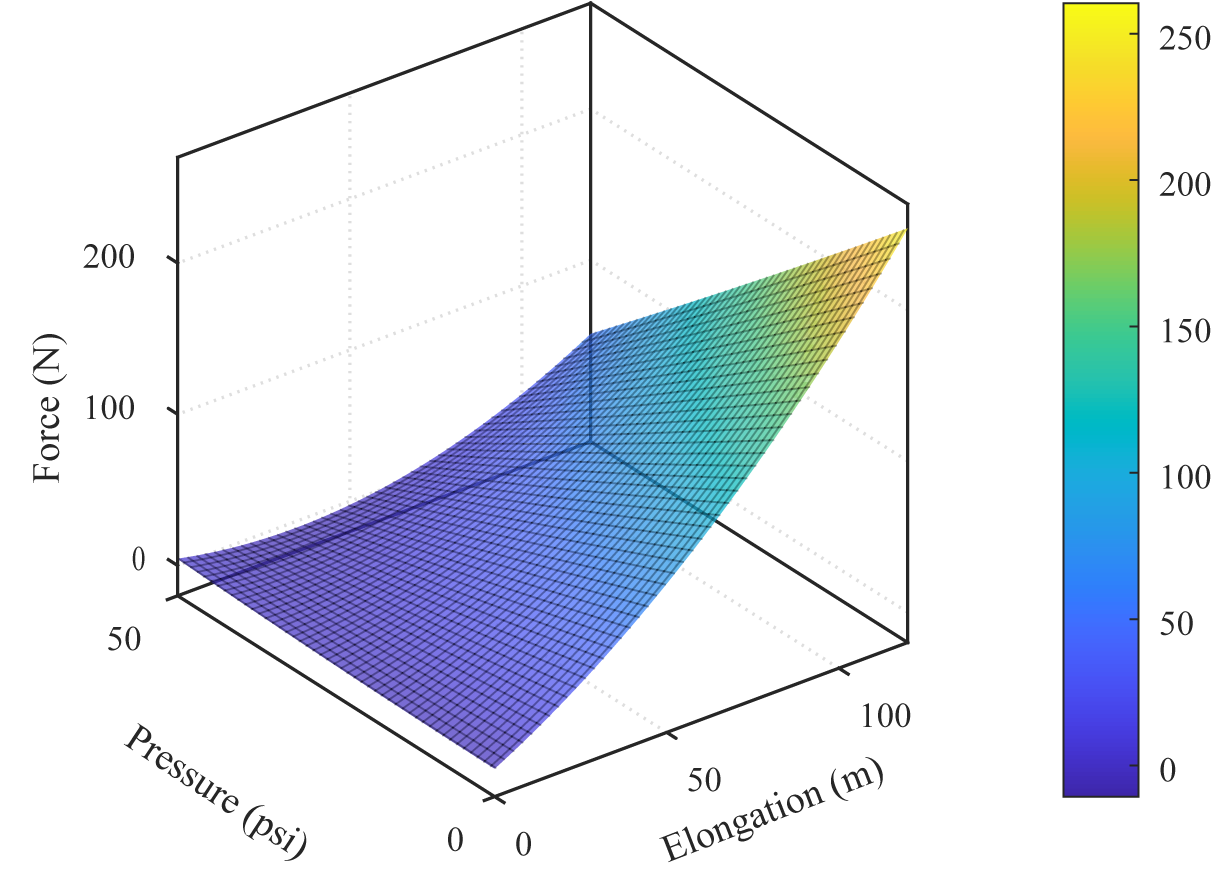}}
\caption{Second-order polynomial surface fit showing IPAM force as a function of elongation and pressure, obtained via bench testing.}
\label{fig: IPAM_model}
\end{figure}

The contraction force of the IPAM depends on both its elongation and internal pressure. To characterize this relationship, we performed a series of controlled tests using a universal testing machine (\#3367, Instron). The IPAM was mounted at its rest length and inflated to 50 psi. The machine then elongated the IPAM to 120 mm, after which the pressure was reduced to a target value between 45 psi and 0 psi in 5 psi increments. After each deflation, the machine returned the IPAM to its rest length. This test sequence was repeated for each target pressure value. During each trial, we recorded internal pressure of the IPAM using a pressure sensor, and measured the elongation and axial force using the Instron machine. The resulting dataset was used to empirically model the IPAM force as a function of pressure and elongation. A second-order polynomial surface was then fit to the data to approximate the active force $F_{\text{active}}$ in terms of pressure $p$ (psi) and elongation $\ell$, as shown in Fig. \ref{fig: IPAM_model}:
\begin{equation}
\begin{aligned}
F_{\text{active}}(\ell, p) &= a + b\ell + cp + d\ell^2 + ep^2 + f\ell p \\
&= -1.4318 + 1.2213\ell + 0.0100p \\
&\quad + 0.0076\ell^2 + 0.0022p^2 - 0.0348\ell p,
\end{aligned}
\label{eq:ipam_force}
\end{equation}
where $F_{\text{active}}$ is the contraction force. This model captures the combined force of the two IPAMs in parallel under all tested combinations of $\ell$ and $p$ within the experimental range.

\subsection{Variable Elastic Actuator Model}

\begin{figure}[t!]
\centerline{\includegraphics[width = 1\linewidth]{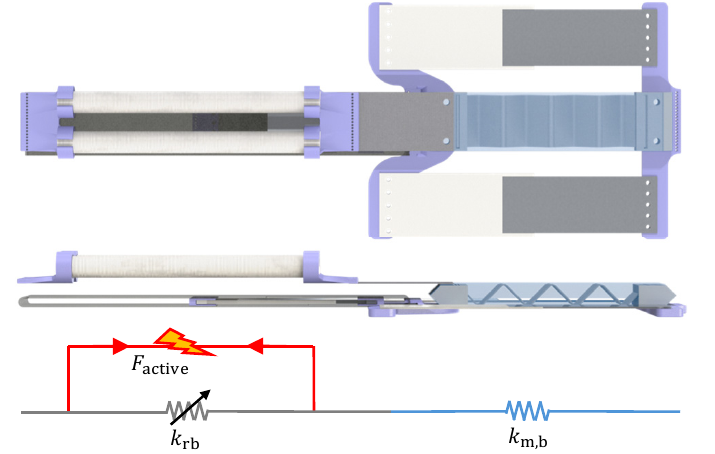}}
\caption{Analytical model of the full device representing total device force as the sum of passive force from the VS resistance band and active force from the IPAM. The origami muscle assembly is in series but is assumed to be inextensible.
}
\label{fig: model_full_device}
\end{figure}

The total axial force output of the full device (shown in Fig. \ref{fig: model_full_device}) is given by:
\begin{equation}
F_{\text{total}}(t) = F_{\text{active}}(\ell(t), p(t)) + k_{\text{rb}}\ell(t),
\label{eq:total_force}
\end{equation}
where $p(t)$ is the internal pressure in psi, $\ell(t)$ is the measured elongation in mm. As shown in Fig. \ref{fig: model_full_device}, the origami muscle assembly is in series with the parallel arrangement of the VS resistance band and the IPAMs. However, for simplicity, we assumed that the stiffness of the origami muscle assembly with the EA clutches engaged ($k_{\text{m,b}}$) was sufficiently large compared to $k_{\text{rb}}$ and can be treated as $\infty$.

\subsubsection{Model-based Device Design}

We first applied the VS resistance band model to select geometric parameters that would yield two desired stiffness levels. Based on prior work (\cite{lamers_feasibility_2018}), we selected 0.8\,N/mm as the base stiffness with the EA clutch disengaged, matching the stiffness of state-of-the-art passive devices. To provide noticeably increased assistance for heavier loads, we targeted a 1.5 times increase in stiffness (1.6\,N/mm) when the EA clutch is engaged. Using Eqn.~\ref{eqn:k-rb}, we solved for the lengths and cross-sectional dimensions of each region of the band to achieve these two target stiffnesses.

For the weight adaptive lifting profiles, the main design goal was to select the IPAM dimensions and the most compact possible pneumatic components to achieve a rapid and adequate increase in the IPAM force at the beginning of the trunk extension phase. First, we obtained the expected IPAM length as a function of time ($\ell(t)$) by measuring the elongation between the back and the thigh attachments during a trunk flexion-extension cycle. We used a wire draw encoder to measure this elongation as described in Sec. \ref{sec:human exp}.A. Further, we identified multiple feasible configurations of IPAM dimensions and pneumatic components. For each of these configurations, we computed the IPAM pressure dynamics using Eqns. \ref{eq:p_dyn_1} or \ref{eq:p_dyn} and, in turn, $F_{\text{active}}(\ell(t), p(t))$ and $F_{\text{total}}(t)$ using Eqns. \ref{eq:ipam_force} and \ref{eq:total_force} respectively. We selected a configuration that was able to provide an $F_{\text{total}}(t)$ of 175\,N within minimum time. We selected the 175\,N target based on the typical peak force of medium force BSDs (\cite{quirk_evaluating_2024}).

\begin{figure*}[th!]
\centerline{\includegraphics[width = 1\linewidth]{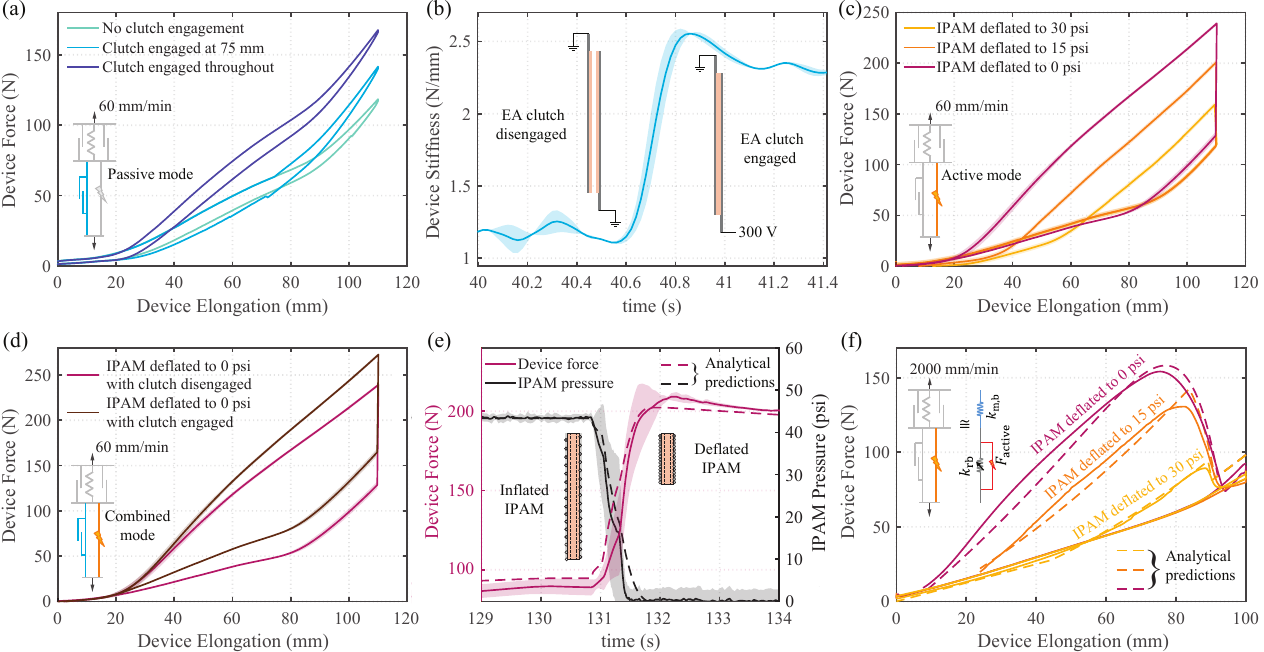}  }
\caption{Characterization results. (a) Force profile tuning in the passive mode. (b) Rapid response of the EA clutch. (c) Force profile tuning in the active mode. (d) Peak device force with combined EA clutch and IPAM actuation. (e) Rapid response of the IPAM and the accurate prediction of the device force by the analytical model. (f) Dynamic force testing in the active mode and the accurate prediction of the dynamic device force by the analytical model. The shaded region represents $\pm1$ standard deviation in all the sub-figures.}
\label{fig: bench test}
\end{figure*}

\section{Device Characterization}\label{sec:Dev Char}

We conducted a series of force tests to characterize the device’s force profile tuning capabilities and validate the analytical model. To isolate the contributions of the VS resistance band and the IPAM, we performed two sets of tests: one in \textit{passive mode} and one in \textit{active mode}. In passive mode, the IPAM remained fully inflated and did not contribute any active force; only stiffness tuning via the EA clutch was evaluated. In active mode, the EA clutch remained disengaged throughout, and the device’s behavior under active force tuning from the IPAM was assessed.

\subsection{Device Characterization in Passive Mode}

To evaluate the EA clutch’s ability to increase the device’s stiffness, we conducted three force tests using a universal testing machine. These tests emulate a typical use case where stiffness is increased during weight-lowering tasks, as depicted in Fig.~\ref{fig: First Figure}(b). We also assessed the EA clutch engagement latency to evaluate its suitability for real-time stiffness tuning.

In each test, the device was elongated to 110\,mm at a constant rate of 60\,mm/min, held for 2\,s at peak elongation, and then returned to its original length at the same rate. In the first test, the EA clutch was disengaged throughout the cycle. In the second, it remained engaged throughout. In the third test, the EA clutch was engaged at 75\,mm during the elongation phase and disengaged again at 75\,mm during the return phase. Each condition was repeated three times.

Fig.~\ref{fig: bench test}(a) shows the resulting force-displacement curves. EA clutch engagement increased the base stiffness from 0.875\,N/mm to 1.313\,N/mm, representing a 50.06\% increase. In the third test, the stiffness transition occurred rapidly once the EA clutch was engaged at 75\,mm, demonstrating the system’s ability to switch stiffness during motion. The stiffness switching latency was measured to be under 300\,ms, as shown in Fig.~\ref{fig: bench test}(b).

Although only one intermediate engagement profile is shown in Fig.~\ref{fig: bench test}(a), additional profiles could be generated by adjusting the EA clutch engagement timing. As expected, the force-displacement curves exhibit hysteresis, with lower forces during unloading than loading. Additionally, all three profiles show a noticeable increase in stiffness beyond 90\,mm of elongation. This behavior results from the fully inflated IPAM being stretched beyond its resting length, which pretensions the inflated IPAM.

\subsection{Device Characterization in Active Mode}

To characterize the IPAM’s ability to modulate active force, we conducted another set of three force tests using the same universal testing machine. These tests simulated the device’s behavior during weight-adaptive lifting assistance, as shown in Fig.~\ref{fig: First Figure}(b). In each test, the device was elongated to 110\,mm at 60\,mm/min with the IPAM fully inflated to 50\,psi. At peak elongation, the IPAM was deflated to a target pressure of 30\,psi in the first test, 15\,psi in the second, and 0\,psi (fully deflated) in the third. The IPAM remained at the target pressure during unloading, and the device was returned to its original length at the same rate. Each condition was repeated three times.

The resulting force profiles, shown in Fig.~\ref{fig: bench test}(c), indicate a clear increase in assistive force as the IPAM deflation level increases. With full deflation to 0\,psi, the actuator added up to 110.71\,N of force at peak elongation. The device also demonstrated the ability to produce a range of intermediate force levels depending on the deflation pressure, supporting its use for weight-adaptive assistance. All three profiles exhibited net energy addition over the flexion-extension cycle, indicating the potential for positive mechanical work on the human body. Although only three force profiles are shown, intermediate profiles could be realized by adjusting the deflation pressure. Further, Fig.~\ref{fig: bench test}(d) shows that the peak force can be increased by 33.32\,N when full IPAM deflation is combined with EA clutch engagement at the start of elongation.

To verify that the IPAM could meet our real-time actuation goals, we assessed its deflation performance. As shown in Fig.~\ref{fig: bench test}(e), the IPAM can add a 103.86\,N of force (79.67\% of its peak) within 800\,ms, satisfying our design requirement. The analytical model used to guide the IPAM and pneumatic system design closely matched the experimental data, accurately capturing the pressure and force dynamics during deflation.

\subsection{Dynamic Characterization}

While the previous tests were conducted under quasi-static conditions (60\,mm/min), real-world back movements occur at faster rates. Thus, we evaluated the device’s performance under dynamic conditions by accounting for two critical factors: IPAM deflation latency and the rate of device elongation during typical lifting.

To simulate real-world device elongation rates, we first measured the elongation velocity of the device during lifting in a simple human experiment. A wire-draw encoder was placed on the thigh attachment, with the wire connected to the back attachment. The encoder data from three object-lifting trials were differentiated to estimate an average elongation rate of approximately 2000\,mm/min. We used this velocity in the dynamic bench test, where the device was elongated by 100\,mm and returned to 0\,mm at 2000\,mm/min. The IPAM was inflated to 50\,psi during elongation and deflated after a 300 ms delay from the start of the return phase. The 300 ms delay in the IPAM deflation onset was included to simulate the expected delay in the detection of trunk extension by our user state estimation algorithm (Sec. \ref{Sec:SM}). We conducted three tests with target deflation pressures of 30\,psi, 15\,psi, and 0\,psi.

Fig.~\ref{fig: bench test}(f) shows that in each case, the IPAM force peaked within the first 25\% of the return phase, demonstrating rapid actuation. The analytical model remained accurate under these dynamic conditions, and despite reduced peak forces due to actuation latency, the device still delivered sufficient force output for lifting assistance. These results confirm that the system can effectively adapt force profiles in real time, even under dynamic conditions.

\section{Optimal Assistive Force Profile}\label{Sec:Moco}

In this section, we show that the lowering and lifting force profiles of our device (Fig. \ref{fig: bench test}(a) and (f)) closely approximate the shape of biomechanically optimal force profiles. We aim to determine the shape of biomechanically optimal force vs RoM profiles for object lowering and lifting. Conventional methods identify the optimal device torque for a given back motion by scaling the biological trunk torque (\cite{poliero_active_2022}). However, this method is not applicable to our device, as it wraps around the user's pelvis, and the moment arm of the device varies highly nonlinearly with the trunk angle. This implies that the shape of the optimal device force profile may be significantly different from the biological trunk torque profile. Thus, a systematic approach to determining optimal force profiles using a coupled device-musculoskeletal model is necessary.

\subsection{Coupled Device-musculoskeletal Model}
\begin{figure}[t]
\centerline{\includegraphics[width = 1\linewidth]{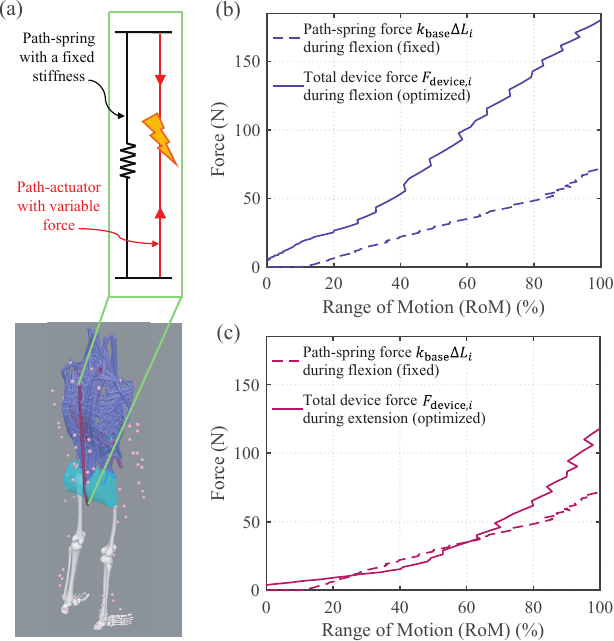}}
\caption{Device force profile optimization for object lowering and lifting. (a) Coupled device-musculoskeletal model. (b) Optimal force profile for object lowering. (c) Optimal force profile for object lifting.}
\label{fig: Mocol}
\end{figure}

For the coupled device-musculoskeletal model, we modified a previously validated OpenSim Lifting Full-Body (LFB) model to incorporate our device (\cite{beaucage-gauvreau_validation_2019, akhavanfar_sharing_2022, akhavanfar_enhanced_2024}). We modeled the base stiffness of the VS resistance band (stiffness with the EA clutch disengaged) using an OpenSim \texttt{PathSpring} with a given stiffness \(k_{\mathrm{base}}\) as shown in Fig. \ref{fig: Mocol}(a). An OpenSim \texttt{PathActuator} with a customizable force \(f_{\mathrm{path}}(t)\) within upper and lower bounds was attached in parallel with the path spring. 
The optimization determined the optimal force profile of this path actuator, separately for weight lowering and lifting.

\begin{figure*}[thpb]
\centerline{\includegraphics[width = 1\linewidth]{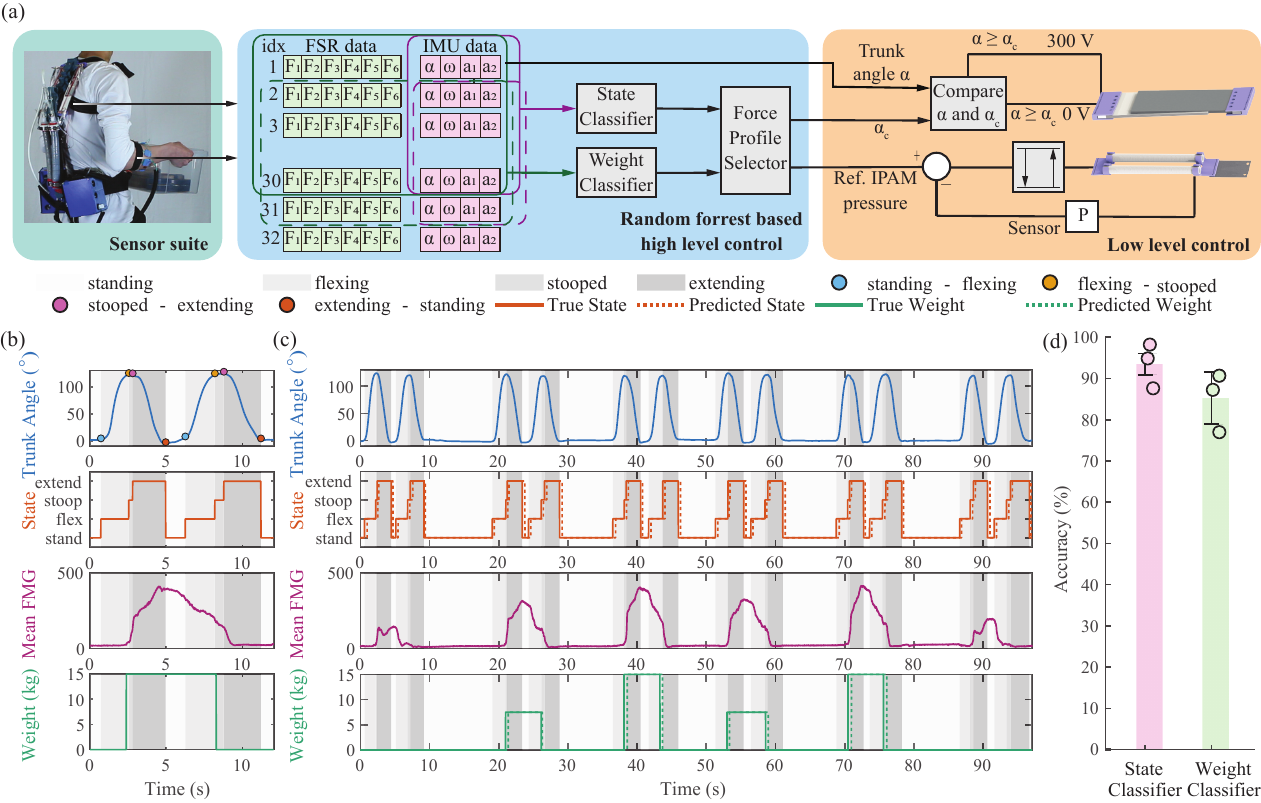}}
\caption{State and weight estimation. (a) Overview of the automated weight adaptive force profile tuning algorithm. (b) Labeling of the training data for the state and the weight classifier. (c) Real-time state and weight classification in a representative subject. (d) Real-time state and weight classification accuracy across three subjects.}
\label{fig: S_M}
\end{figure*}

\subsection{Optimization Problem Formulation}
We used open-source motion-capture and force-plate data of a human subject lowering and lifting a weight from Ref.~(\cite{beaucage-gauvreau_validation_2019}) and used OpenSim's Static Optimization (SO) tool to identify device force profiles that reduce biological muscle effort. SO solves, at each discrete time frame \(i\), a muscle activation-to-force inverse-dynamics problem that minimizes a muscle activation-based cost while satisfying joint-moment equilibrium.

At each frame \(i\) we solve the following equilibrium constraint (for each biological joint \(j\)):
\begin{multline}
    \sum_{m=1}^{M} \big[ a_{m,i}\, f_m(F_m^{0},\ell_{m,i},v_{m,i})\big]\, r_{m,j,i}
\\ + r_{\mathrm{device},j,i}\,F_{\mathrm{device},i} = \tau_{j,i},
\label{eq:equilibrium_flv}
\end{multline}
while minimizing the activation-based objective
\begin{equation}
J_i \;=\; \sum_{m=1}^{M} \, a_{m,i}^2 \;+\; \, a_{\mathrm{path},i}^2,
\label{eq:objective_with_path_flv}
\end{equation}
where \(a_{m,i}\in[0.01,1]\) is the activation of muscle \(m\) at frame \(i\). The scalar \(a_{\mathrm{path},i}\in[0.001,0.01]\) is the (unitless) activation-like control of the OpenSim \texttt{PathActuator} at frame \(i\) (we chose the small range for \(a_{\mathrm{path}}\) so the optimizer can reach large physical actuator forces with small \(a_{\mathrm{path}}\) values and thereby preferentially use the actuator). In Eqn. \ref{eq:equilibrium_flv}, \(M\) is the number of muscles, \(F_m^{0}\) is the maximum isometric force of muscle \(m\), and $f_m(F_m^{0},\ell_{m,i},v_{m,i})$ is the active-fiber force scaling (force–length–velocity) computed by the OpenSim muscle model at fiber length \(\ell_{m,i}\) and shortening velocity \(v_{m,i}\). The scalar \(r_{m,j,i}\) is the moment arm of muscle \(m\) about joint \(j\) at frame \(i\) (collectively \(\mathbf{R}_i\)), and \(\tau_{j,i}\) are the generalized joint torques from inverse dynamics (collectively \(\boldsymbol{\tau}_{\mathrm{req},i}\)). The device moment arm about joint \(j\) is \(r_{\mathrm{device},j,i}\) (collectively \(\mathbf{r}_{\mathrm{device},i}\)), and the total device force at frame \(i\) is $F_{\mathrm{device},i} \;=\; k_{\mathrm{base}}\,\Delta L_i \;+\; f_{\mathrm{path},i}$ where \(\Delta L_i = L_i - L_0\) is the instantaneous device elongation (with \(L_i\) the current path length and \(L_0\) the rest length) and \(k_{\mathrm{base}}\) is the \texttt{PathSpring} stiffness. The \texttt{PathActuator} force is given by $f_{\mathrm{path},i} = a_{\mathrm{path},i}\, F_{\mathrm{path}}^{\max}$ with \(F_{\mathrm{path}}^{\max}\) it's maximum force. We selected \(a_{\mathrm{path}}^{\min}=0.001\), \(a_{\mathrm{path}}^{\max}=0.01\) and chose \(F_{\mathrm{path}}^{\max}\) so that \(a_{\mathrm{path}}^{\max} F_{\mathrm{path}}^{\max} \approx F_{\mathrm{phys}}^{\max}\) (the physical IPAM peak).

We ran SO to obtain an optimized total device force $F_{\mathrm{device},i}$ separately for weight lowering and lifting. For lowering, we ran SO only during the trunk flexion phase of a trunk flexion-extension cycle. For lifting, we ran SO only during the trunk extension phase.

\subsection{Optimization Results}
The resultant optimal force profiles for weight lowering and lifting are shown in Fig. \ref{fig: Mocol}(b) and (c), respectively. The passive force applied by the fixed stiffness \texttt{PathSpring} during trunk flexion is also shown. The optimal total device force for lowering is approximately linear with a higher stiffness as compared to the base stiffness of the device. Thus, the shape of the optimal lowering profile is similar to the lowering profile obtained by increasing the stiffness of the device for a heavier weight (Fig. \ref{fig: bench test}(a)). The optimal total device force for lifting appears to add an instantaneous active force to the passive force at the beginning of the extension phase (RoM = 100\%), similar in shape to our device's lifting force profile (Fig. \ref{fig: bench test}(d)). In summary, both the lowering and lifting force profiles implemented in our device closely mimic the biomechanically optimal force profiles.

\section{State and Weight Estimation}\label{Sec:SM}

This section describes the state and weight classification algorithm used to enable automated tuning of assistive force profiles during lifting and lowering tasks, as shown in Fig.~\ref{fig: S_M}(a). The algorithm uses data from an upper-back IMU and an FMG armband.

The state classifier segments a user's motion into one of four discrete trunk states: standing, flexing, stooped, or extending. It uses four IMU features: sagittal-plane $\upalpha$ trunk angle, trunk angular velocity $\upomega$, and two orthogonal components of total linear acceleration in the sagittal plane $\text{a}_1$ and $\text{a}_2$. The weight classifier classifies the object being handled into one of three weight categories: 0\,kg, 7.5\,kg, or 15\,kg. It uses both IMU and FMG data (FSR readings $\text{F}_\text{1}$ - $\text{F}_\text{6}$) as input.

Both classifiers were implemented using random forest models trained on lab-collected data. The classifiers' outputs were passed to a force profile selector that triggered appropriate device behavior: weight-adaptive lowering or lifting assistance. The force profile selector identifies the start of a lowering phase as a transition from standing to flexing and the start of a lifting phase as a transition from stooped to extending. The weight classification output then determines the appropriate trunk angle $\upalpha_{\text{c}}$ (\% RoM) for EA clutch engagement and reference IPAM pressure. For weightlifting, the IPAM deflates at the beginning of trunk extension to 30\,psi, 15\,psi, or 0\,psi for 0\,kg, 7.5\,kg, or 15\,kg, respectively. For weight lowering, the EA clutch remained disengaged for 0\,kg, engaged at midpoint of RoM for 7.5\,kg, and engaged at onset of RoM for 15\,kg.

\subsection{State and Weight Classifier}
The state classifier used a 0.3\,s window (30 samples at 100\,Hz) of IMU data as input, forming a 30$\times$4 matrix. From this, 32 time-domain summary features were extracted, including mean, standard deviation, minimum, maximum, start, end, slope, and delta mean for each channel. All feature vectors were z-score normalized. Window size and feature set were chosen based on preliminary testing across multiple configurations (0.1\,s to 0.5\,s windows and various feature subsets). A 0.3\,s window with the selected features achieved an average classification accuracy above our target of 90\%, making it the final configuration.

The weight classifier used the same 0.3\,s window duration and processed both IMU and FMG data (30$\times$10). From this, 80 summary features were extracted using the same feature types as the state classifier. Feature vectors were z-score normalized. Similar to the state classifier, window size and feature set were selected through preliminary optimization.

\subsection{Training Data Collection}
Subject-specific models were trained for both the state and weight classifiers. The training data were collected in our laboratory from three participants during object lifting experiments. This dataset was collected without the device.

Each subject performed 30 trials: 10 trials each with 0\,kg, 7.5\,kg, and 15\,kg weights, randomized in order. Each trial included two complete flexion-extension cycles: (1) a lifting cycle—flexion from standing to lifting the weight and returning to the standing position with the weight, and (2) a lowering cycle—flexion to lower the weight and return to the standing position without it. This resulted in 10 lifting and 10 lowering cycles per weight class per subject.

\subsection{Data Pre-processing and Labeling}
No additional filtering was applied to the IMU or FMG signals. Cycle start and end points were manually labeled by visually inspecting time-synchronized plots of trunk angle, trunk angular velocity, and video recordings, as shown in Fig.~\ref{fig: S_M}(b).

Using these points, the state labels were generated as follows: standing (from previous extension end to flexion start), flexing (from flexion start to end), stooped (from flexion end to extension start), and extending (from extension start to end). The weight labels were assigned by observing the FMG signal along with the synchronized video for each cycle.

\subsection{Real-time Performance and Limitations}
In real-time inference, classifier outputs were smoothed using a 10-sample dwell-time filter to reduce state toggling. As shown in Fig.~\ref{fig: S_M}(c), the state classifier closely tracks true motion states labeled offline, and the weight classifier accurately identifies the handled weight. Additional real-time results are presented in the supplementary movie. Across the three subjects, the average real-time classification accuracy was $93.47 \pm 2.59\%$ for state prediction and $85.23 \pm 6.31\%$ for weight prediction (Fig.~\ref{fig: S_M}(d)).

Currently, both state and weight classifiers are subject-specific, which limits generalizability across users. While our approach enables high per-user accuracy, it introduces calibration overhead and reduces scalability. Future work will explore subject-independent models.

\section{Human Experiments}\label{sec:human exp}

We conducted lifting and lowering experiments with human subjects to validate (1) the on-body tuning abilities and (2) the biomechanical benefits of the device. All experimental protocols were approved by the Institutional Review Board at Arizona State University (STUDY00020698).

\subsection{On-body Device Characterization}
We first evaluated the implementation of the force profiles described in Sec.~\ref{sec:Dev Char} during object lifting and lowering on the human body.

\begin{figure*}[tb]
\centerline{\includegraphics[width = 1\linewidth]{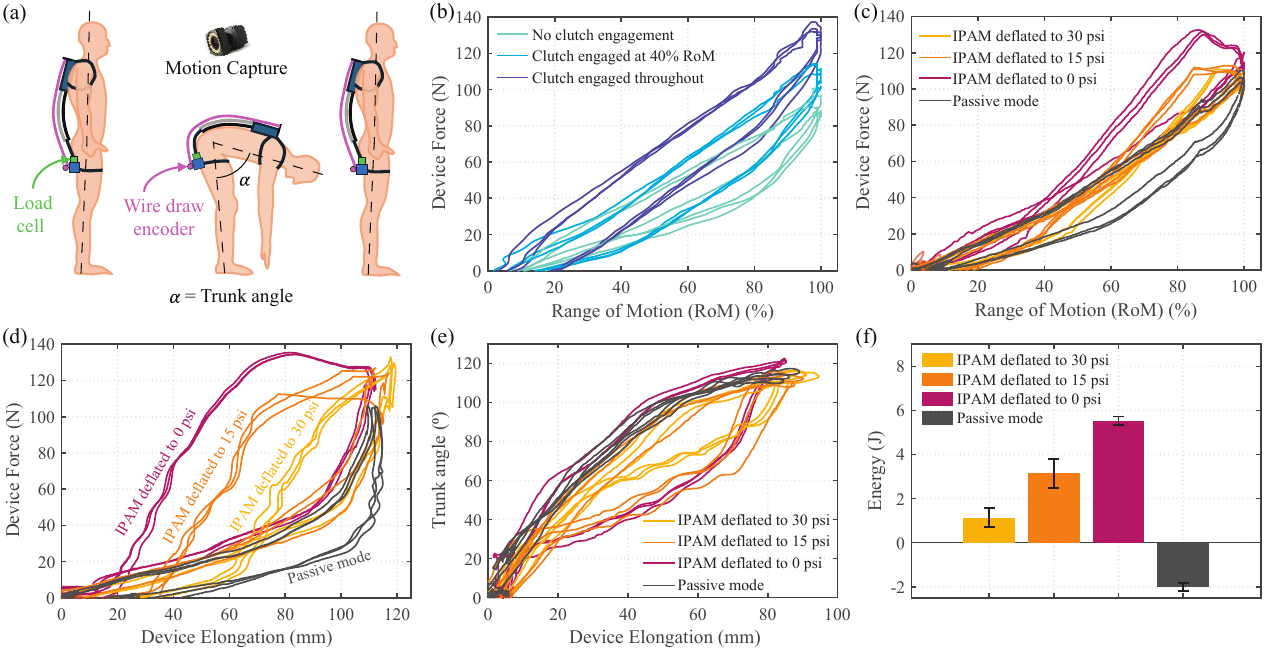}}
\caption{On-body Device Characterization. (a) Experimental protocol: Motion capture used to determine the trunk angle and the RoM, load cell used to measure the device force, and the wire draw encoder used to measure device elongation. (b) Force profile tuning in the passive mode that facilitates weight-adaptive lowering assistance. (c) Force profile tuning in the active mode that facilitates weight-adaptive lifting assistance. (d) Force vs device elongation in the active mode. (e) Trunk angle vs device elongation in the active mode. (f) Increasing energy added to the human body over a flexion-extension cycle with the increasing active force level.}
\label{fig: human experiment}
\end{figure*}

\subsubsection{Experimental setup}
To measure device force, a load cell (Model LCM300, FUTEK; accuracy $\pm 0.25\%$) was placed between the bottom of the resistance band and the thigh attachment, as shown in Fig.~\ref{fig: human experiment}(a). Device elongation was recorded using a wire-draw encoder mounted on the thigh attachment, with the wire connected to the back attachment. These sensors were integrated into the device only for on-body characterization and were not part of the device. The trunk angle $\alpha$ was measured using an 8-camera motion capture system (Vicon Motion Systems Ltd., Oxford, UK) with the Vicon Plug-in Gait full-body marker set~(\cite{nexus_full_2024}).

\subsubsection{Experimental protocol and data processing}
Three participants (height: $171.3\pm 0.9$ cm, weight: $53.87\pm 3.6$ kg, age: $22.0\pm 0.82$ years) were recruited for the on-body characterization. After the study protocol was explained, participants wore the device, load cell, and wire-draw encoder, followed by the placement of reflective markers. A 30-second standing calibration trial was recorded prior to data collection.

Each participant performed object lifting trials under three weight conditions: 0\,kg (empty crate), 7.5\,kg, and 15\,kg. Each trial consisted of two flexion-extension cycles: (1) a lifting cycle, which began from standing, included flexion to pick up the weight, and ended after returning to standing with the weight; and (2) a lowering cycle, which began with flexion to lower the weight and ended after returning to standing without the weight.

During the lifting cycle, the device applied one of three active force profiles (similar to Fig.~\ref{fig: bench test}(f)). The IPAM was inflated during flexion and deflated at the start of extension, based on the lifted weight: 30\,psi for 0\,kg, 15\,psi for 7.5\,kg, and 0\,psi for 15\,kg. During the lowering cycle, the device applied one of three passive profiles (similar to Fig.~\ref{fig: bench test}(a)). The EA clutch remained disengaged for 0\,kg, engaged at the midpoint of RoM for 7.5\,kg, and engaged at the start of RoM for 15\,kg. Profile selection during both lifting and lowering was controlled using our real-time state and weight estimation algorithm described in Sec.~\ref{Sec:SM}.

Kinematic data were filtered using a fourth-order, zero-lag low-pass filter at 6\,Hz. Raw load-cell and encoder data were used without filtering. Trunk angle was computed by summing the thorax and hip angles relative to the absolute pelvis, following Vicon Plug-in Gait conventions.

\subsubsection{Results}
Fig.~\ref{fig: human experiment}(b) shows device force versus trunk angle for the three weight conditions in passive mode during lowering. Compared to bench testing, greater hysteresis was observed, likely due to deformation and compliance at the device-body interface.

Fig.~\ref{fig: human experiment}(c) shows force versus trunk angle during lifting in active mode. The force addition increased with the reduction of the IPAM deflation pressure target. However, the separation between force profiles was less pronounced than in the bench test (Fig.~\ref{fig: bench test}(f)). When force was plotted against device elongation (Fig.~\ref{fig: human experiment}(d)), the profiles closely matched the bench-test trends. In order to investigate further, we evaluated the trunk angle vs device elongation plot shown in Fig. ~\ref{fig: human experiment}(e). Higher IPAM pressures produced a more rapid reduction in device length at the onset of extension. This behavior may be due to two factors: (1) local body deformation at contact regions under higher forces, and (2) altered posture, such as reduced spinal curvature, in response to the assistive force. Both effects could reduce device elongation without a corresponding change in trunk angle, explaining the separation between force profiles in Fig.~\ref{fig: human experiment}(c) being less pronounced than in Fig.~\ref{fig: bench test}(f).

\begin{figure}[b!]
\centerline{\includegraphics[width = 1\linewidth]{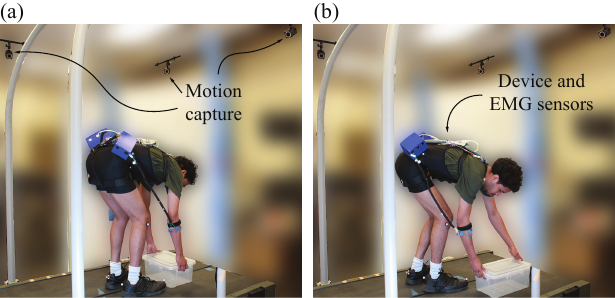}}
\caption{(a) Experimental setup and protocol for (a) the symmetric lifting/lowering task. (b) the asymmetric lifting/lowering task.}
\label{fig: emg experiment protocol}
\end{figure}

\begin{figure*}[htpb]
\centerline{\includegraphics[width = 1\linewidth]{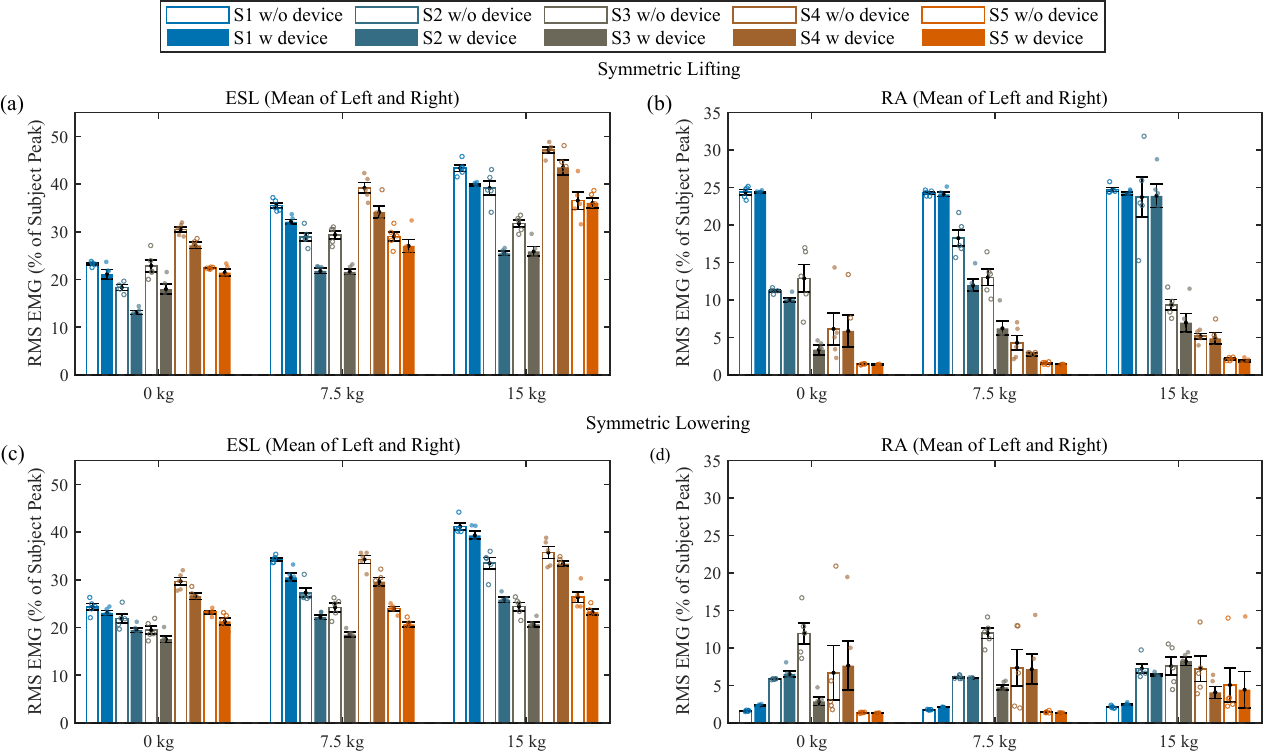}}
\caption{Evaluation of back extensor and trunk flexor activity during symmetric object lifting and lowering. (a) Effect on the back extensors during weightlifting: The peak normalized root mean squared (RMS) EMG of ESL (mean of left and right) for individual subjects during weightlifting without and with the device. (b) Effect on the trunk flexors during weightlifting: The peak normalized RMS EMG of RA (mean of left and right) for individual subjects during weightlifting without and with the device. (c) Effect on the back extensors during weight lowering: The peak normalized RMS EMG of ESL (mean of left and right) for individual subjects during weight lowering without and with the device. (d) Effect on the trunk flexors during weight lowering: The peak normalized RMS EMG of RA (mean of left and right) for individual subjects during weight lowering without and with the device.}
\label{fig: emg experiment}
\end{figure*}

Peak forces recorded during human trials were generally lower than those measured during bench testing, likely due to frictional losses and compliance at the device-body interface. Fig.~\ref{fig: human experiment}(f) quantifies the energy added by the device to the human body over flexion-extension cycles across the three subjects. This energy was computed as the area under the device force vs device elongation curve over a flexion-extension cycle. As expected, energy addition was negative in the passive mode due to hysteresis. In contrast, the active mode produced net positive energy input, validating the benefit of including an active element in parallel with a passive element.

\subsection{Muscle Activity Analyses}
We evaluated the effect of the device on muscle activation during symmetric and asymmetric lifting and lowering tasks. Surface EMG measurements were used to assess activation of both back extensor and trunk flexor muscles.

\subsubsection{Experimental setup and protocol}
As shown in Fig.~\ref{fig: emg experiment protocol}, the experimental setup was similar to that used for the on-body device characterization. Trunk angle was recorded using an 8-camera motion capture system (Vicon Motion Systems Ltd., Oxford, UK) with the Plug-in Gait full-body marker set~(\cite{nexus_full_2024}). EMG signals were recorded using wireless surface EMG sensors (Trigno Avanti, Delsys Inc).

For the symmetric lifting and lowering task, bilateral EMG data were collected from the Erector Spinae Longissimus (ESL; back extensors) and Rectus Abdominis (RA; trunk flexors). For the asymmetric tasks, bilateral EMG was recorded from the ESL and External Obliques (EO; trunk rotators).

Five participants (S1-S5, height: $173.6\pm 3.14$ cm, weight: $66.28\pm 18.11$ kg, age: $22.4\pm 2.06$ years) completed the symmetric lifting and lowering protocol. After providing informed consent, participants were instrumented with the assistive device, EMG sensors, and motion capture markers. A 30-second standing trial was recorded for motion capture calibration.

Each subject performed 5 trials per experimental condition. Each trial consisted of two flexion-extension cycles: one for lifting the object and one for lowering it (as defined in Sec. \ref{sec:human exp}.A). Subjects performed trials with 0 kg, 7.5 kg, and 15 kg weights. A metronome ensured consistent cycle timing (6 s per trial), and weight order was randomized across subjects to minimize potential ordering effects. During lifting, the device applied weight-adaptive active profiles; during lowering, it applied weight-adaptive passive profiles, consistent with the on-body characterization trials.

\begin{figure}[t!]
\centerline{\includegraphics[width = 1\linewidth]{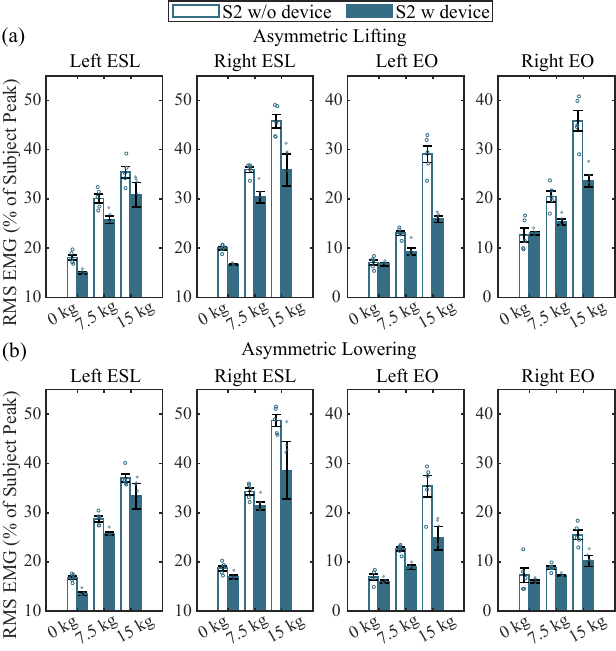}}
\caption{Evaluation of back extensor and trunk flexor activity during asymmetric object lifting and lowering (a) Reduction in the peak normalized RMS EMG of left ESL, right ESL, left EO, and right EO during asymmetric weight lifting in a single subject with the device. (b) Reduction in the peak normalized RMS EMG of left ESL, right ESL, left EO, and right EO during asymmetric weight lowering in a single subject with the device.}
\label{fig: emg experiment asym}
\end{figure}

\subsubsection{Data processing}
Raw EMG signals were band-pass filtered (20--450\,Hz) using a fourth-order zero-lag Butterworth filter, rectified, and then low-pass filtered at 6\,Hz to extract the EMG envelope. The resulting envelope was segmented into individual cycles and normalized in time. To compare muscle activity with and without the device, we computed peak-normalized RMS EMG, defined as the RMS EMG of a muscle during a given trial divided by the peak EMG value observed across all trials for that subject.

\subsubsection{Results}
Fig.~\ref{fig: emg experiment}(a) shows a consistent reduction in ESL activation across all weights and all five subjects during lifting. On average, peak-normalized ESL activation decreased by $15.0 \pm 10.1\%$ with the device across the subjects and the weights. In comparison, RA activation (Fig.~\ref{fig: emg experiment}(b)) was also reduced by $17.8 \pm 22.4\%$ on average across the subjects and the weights, although the pattern of reduction was more variable. The reduction in ESL activity is within the range reported for state-of-the-art active devices (\cite{kermavnar_effects_2021}), though on the lower end of that range. This is likely due to the semi-active nature of our device, where users still perform mechanical work to elongate the passive element before receiving active assistance. Given that our device is significantly lighter than most fully active systems (Table~\ref {table:g}), this moderate EMG reduction may represent an acceptable trade-off.

During the lowering phase, peak-normalized ESL activation was similarly reduced across all subjects and weights (Fig.~\ref{fig: emg experiment}(c)), with an average decrease of $12.4 \pm 5.8\%$ across the subjects and the weights. RA activation (Fig.~\ref{fig: emg experiment}(d)) showed no adverse increases and was reduced on average by $6.8 \pm 31.6\%$ across the subjects and the weights.

To explore the feasibility of assisting asymmetric trunk motion, we conducted an exploratory trial with one subject. As shown in Fig.~\ref{fig: emg experiment asym} (a), the device reduced RMS EMG in both the left and right ESL during asymmetric lifting by $14.8 \pm 2.2\%$ and $18.1 \pm 3.2\%$ across the weights, respectively. Reductions were also observed in the left and right EO muscles ($26.1 \pm 20.6\%$ and $19.1 \pm 19.1\%$ across the weights, respectively), which contribute to trunk rotation. During lowering (Fig.~\ref{fig: emg experiment asym} (b)), the device reduced RMS EMG in both the left and right ESL during asymmetric lifting by $13.22 \pm 5.47\%$ and $12.99 \pm 6.76\%$ across the weights, respectively. Reductions were also observed in the left and right EO muscles ($28.06 \pm 14.06\%$ and $23.06 \pm 9.71\%$ across the weights, respectively). These findings suggest that the soft, body-conforming design of the device may also enable effective assistance during asymmetric tasks. However, because these results are based on a single subject, they should be considered preliminary, and we refrain from generalizing to broader populations.

\section{Conclusion}\label{sec:conclusion}
We presented a lightweight (1.97 kg), compact, and soft BSD that enables tunable assistive force profiles through a parallel combination of a passive element and an active element. The proposed design advances beyond existing soft passive BSDs by enabling force tuning without the added bulk and weight typically associated with fully active systems. An accurate analytical model was developed to support device design for implementing desired force profiles. Further, we demonstrated real-time weight-adaptive force profile tuning during lifting and lowering, facilitated by an automated weight detection algorithm. Finally, EMG analyses with five human subjects confirmed that the device effectively reduces back extensor muscle activity during lifting and lowering tasks, validating its biomechanical benefits. We anticipate that BSDs employing our tuning strategy have potential for real-world applications, particularly for assisting older adults and individuals with reduced trunk muscle strength in daily living scenarios.

\bibliographystyle{SageH} 
\bibliography{references}

\end{document}